\documentstyle[12pt]{article}
\begin{document}

\def\a{\alpha}
\def\b{\beta}
\def\ch{\chi}
\def\d{\delta}
\def\e{\epsilon}
\def\f{\phi}
\def\g{\gamma}
\def\h{\eta}
\def\i{\iota}
\def\j{\psi}
\def\k{\kappa}
\def\l{\lambda}
\def\m{\mu}
\def\n{\nu}
\def\o{\omega}
\def\p{\pi}
\def\q{\theta}
\def\r{\rho}
\def\s{\sigma}
\def\t{\tau}
\def\u{\upsilon}
\def\x{\xi}
\def\z{\zeta}
\def\D{\Delta}
\def\F{\Phi}
\def\G{\Gamma}
\def\J{\Psi}
\def\L{\Lambda}
\def\O{\Omega}
\def\P{\Pi}
\def\S{\Sigma}
\def\U{\Upsilon}
\def\X{\Xi}
\def\T{\Theta}

\def\Ab{\bar{A}}
\def\gi{g^{-1}}
\def\li{{ 1 \over \l } }
\def\zb{\bar{z}}
\def\Tb{\bar{T}}
\def\pp{ \partial }
\def\pb{\bar{\partial }}
\def\be{\begin{equation}}
\def\ee{\end{equation}}
\def\ben{\begin{eqnarray}}
\def\een{\end{eqnarray}}

\addtolength{\topmargin}{-0.8in}
\addtolength{\textheight}{1in}
\hsize=16.5truecm
\hoffset=-.6in
\baselineskip=7mm

\thispagestyle{empty}
\begin{flushright} \ Aug. \ 1994\\
SNUCTP 94-83\\
\end{flushright}

\begin{center}
 {\large\bf Deformed Minimal Models and Generalized Toda Theory }\\[.1in]
\vglue .5in
 Q-Han Park \footnote{ E-mail address; qpark@nms.kyunghee.ac.kr }
 ~  and ~
 H. J. Shin \footnote{ E-mail address; hjshin@nms.kyunghee.ac.kr }\\
\vglue .5in
{\it Department of Physics \\
and \\
Research Institute for Basic Sciences \\
 Kyunghee University\\
Seoul, 130-701, Korea}
\\[.5in]
{\bf ABSTRACT}\\[.2in]
\end{center}
\vglue .1in
We introduce a generalization of $A_{r}$-type Toda theory based on a
non-abelian group G, which we call the $(A_{r},G)$-Toda theory, and its affine
extensions in terms of gauged Wess-Zumino-Witten actions with deformation
terms. In particular, the affine $(A_{1},SU(2))$-Toda theory describes the
integrable deformation of the minimal conformal theory for the critical Ising
model by the operator $\Phi_{(2,1)}$. We derive infinite conserved charges and
soliton solutions from the Lax pair of the affine $(A_{1}, SU(2))$-Toda theory.
Another type of integrable deformation which accounts for the
$\Phi_{(3,1)}$-deformation of the minimal model is also found in the gauged
Wess-Zumino-Witten context and its infinite conserved charges are given.
\newpage

Recently, using the language of operator algebra, Zamolodchikov has shown that
there exist some relevant perturbation around conformal field theory  which
preserve integrability.[1] In particular, when degenerate fields $\F_{(1,2)},
\F_{(2,1)}, \F_{(1,3)}$ and $\F_{(3,1)}$ are taken as the perturbations, he
suggested that the resulting field theories may possess non-trivial integrals
of motion and worked out explicitly for several examples.  In the Lagrangian
framework, there have been attempts to explain these particular perturbations
in terms of the  affine extension of Toda field theories[2][3].
In general, arbitrary coset conformal models can be  formulated  in terms of
the gauged Wess-Zumino-Witten (WZW) action[4].  Using this fact and also
generalizing the recent work of Bakas for the parafermion coset model[11],  one
of us (Q.P.) has recently shown that an integrable deformation of G/H-coset
models is possible when the gauged WZW action for the G/H-coset model is added
by  a potential energy term $Tr(gTg^{-1}\Tb )$, where algebra elements $T, \Tb
$ belong to the center of the algebra {\bf h} associated with the subgroup
H[6].

 In this Letter, we consider two  types of integrable deformations, by
operators $\F_{(2,1)}$ and $\F_{(3,1)}$, of the minimal model corresponding to
the coset $(SU(2)_{N} \times SU(2)_{N}) /SU(2)_{2N}$ where $N$ denotes the
level. This corresponds to the deformation of the critical Ising model for
$N=1$ and  that of $c=1$ theory in the super conformal minimal series for
$N=2$.  We formulate the minimal model in terms of the gauged
Wess-Zumino-Witten(WZW) action and show that  the integrable deformations by
$\F_{(2,1)}$ and $\F_{(3,1)}$  can be obtained by adding  potential terms;
Tr$(g^{-1}_{1}g_{2} + g^{-1}_{2}g_{1})$ and $\mbox{Tr}(g_1^{-1} L^a g_1 L^b)
\mbox{Tr}(g_2^{-1} M^a g_2 M^b)$ respectively, where $g_{1}, g_{2}$ are Lie
group $SU(2)$-valued fields and  $\{ L^{a} \} ,
\{ M^{a} \} $ are two sets of generators of the Lie algebra $su(2)$. In
particular, the action for the $\F_{(2,1)}$-deformation suggests a natural
generalization of the abelian $A_{r}$-type Toda theory to the non-abelian
$(A_{r}, G)$-Toda theory for a non-abelain group $G$ and its affine extensions
whereas the action for the $\F_{(2,1)}$-deformation itself becomes the affine
$(A_{1}, SU(2))$-Toda theory. We demonstrate  the integrability of both
deformed models by deriving Lax pairs for them and also from which infinitely
many conserved charges. We also derive  n-soliton solutions  for the affine
$(A_{1}, SU(2))$-Toda theory.

Recall that a lagrangian of the G/H-coset model is given in terms of the gauged
 WZW functional[4], which in light-cone variables is
\be
S(g,A, \Ab ) = S_{WZW}(g) +
{1 \over 2\pi }\int \mbox {Tr} (- A\pb g \gi + \Ab \gi \pp g
 + Ag\Ab \gi - A\Ab )
\ee
where $S_{WZW}(g)$ is the usual WZW action [5] for a map $g  :  M \rightarrow
$G on two-dimensional Minkowski space $M$.  The connection $A, \Ab$ gauge the
anomaly free subgroup H of G. In this Letter, we take the diagonal
embedding of H in $G_{L} \times G_{R}$, where $G_{L}$ and $G_{R}$ denote left
and right group actions by multiplication $(g \rightarrow g_{L}gg_{R}^{-1})$,
so that Eq.(1) becomes invariant under the vector gauge transformation $(g
\rightarrow hgh^{-1}$ with $h :  M \rightarrow $H). The restriction to  the
coset  $(SU(2)_{N} \times SU(2)_{N}) / SU(2)_{2N}$ where $N$ denotes the level
of the Kac-Moody algebra is defined by the functional
$ \int [dg_{1}][dg_{2}][dA][d\Ab ] exp(iI_{0}(g_{1}, g_{2}, A , \Ab ))$ where
\be
 I_{0}(g_{1}, g_{2}, A , \Ab ) =  NS_{WZW}(g_{1}, A, \Ab ) + NS_{WZW}(g_{2}, A,
\Ab  )
\ee
where $A$ and $\Ab  $ gauge simultaneously the diagonal subgroups of $SU(2)
\times SU(2)$.  The classical equations of motion for $g_{1}$ and $g_{2}$ arise
in a form of  zero curvature condition,
\ben
&& [ \ \pp + \gi_1 \pp g_1 + \gi_1 A g_1 \ , \ \pb + \Ab \ ] = 0 \nonumber \\
&& [ \ \pp + \gi_2 \pp g_2 + \gi_2 A g_2 \ , \ \pb + \Ab \ ] = 0
\een
whereas variations with respect to  $A$ and $\Ab $ give the constraint
equation,
\ben
 - \pb g_{1} \gi_{1} &+& g_{1}\Ab \gi_{1} - \pb g_{2} \gi_{2} + g_{2}\Ab
\gi_{2} - 2\Ab  = 0
\nonumber \\
 \gi_{1} \pp g_{1}  &+& \gi_{1} A g_{1} +  \gi_{2} \pp g_{2}  +
\gi_{2} A g_{2}- 2A = 0 \ .
\een
\\

{\bf $\Phi_{(2,1)}$-deformation and generalized Toda theory }
\\

Having introduced an action for the coset model, we now assert that an
integrable deformation is possible when we add to the action a potential term
in the following way:
\be
I(g_{1},g_{2},A,\Ab ,\k ) = I_{0}(g_{1},g_{2},A,\Ab ) - {N\k \over 2\pi }\int
\mbox{Tr} (g^{-1}_{1}g_{2} + g^{-1}_{2}g_{1}) ,
\ee
where $\k $ is a coupling constant. The potential term transforms at the
classical level as (doublet, singlet) in the convention of coset conformal
field theory so that it corresponds to the operator $\Phi_{(2,1)}$. This
changes Eq.(3) by
\ben
&& [\  \pp + \gi_{1} \pp g_{1} + \gi_{1} A g_{1} \ , \ \pb + \Ab \ ] - \k (
g^{-1}_{1}g_{2} - g^{-1}_{2}g_{1})  =   0  \nonumber \\
&& [\  \pp + \gi_{2} \pp g_{2} + \gi_{2} A g_{2} \ , \ \pb + \Ab \ ]  + \k
(g^{-1}_{1}g_{2} - g^{-1}_{2}g_{1})  = 0
\een
while leaving the constraint equation unchanged. The main observation in
proving the integrability of the model is that  Eq.(6) is precisely the
integrability condition of the linear $4 \times 4$ matrix equations with a
spectral parameter $\l $,
\be
L_{1}(\l)\Psi \equiv [ \pp +  U_0 - \l T
]\Psi = 0 \  \ ; \  \  L_{2}(\l )\Psi \equiv (\pb +{\bar {\cal A}}+ \li V_1
)\Psi = 0
\ee
where
\be  U_0 = {\cal G}^{-1} \pp {\cal G} + {\cal G}^{-1}{\cal A}{\cal G} \ , \ V_1
= {\cal G}^{-1}\Tb {\cal G} \ , \  T = i\k \S \ ,  \ \Tb = i\S
\ee
 and
\be
{\cal G}  \equiv \left( \begin{array}{ccc}
g_1 & ~ & 0  \\ {}~           & ~   &  ~ \\
 0  & ~ & g_2
          \end{array}   \right) \ , \
 {\cal A} \equiv \left( \begin{array}{ccc}
A & ~ & 0  \\ {} ~          &  ~  &   \\
 0  & ~ & A
          \end{array}   \right) \ , \
 {\bar {\cal A}} \equiv \left( \begin{array}{ccc}
\bar{A} & ~ & 0  \\ {}~           & ~   &  ~ \\
 0  & ~ & \bar{A}
          \end{array}   \right) \ , \
\S \equiv \left( \begin{array}{ccc}
0  & ~ & 1  \\ {}~           & ~   &  ~ \\
 1  & ~ & 0
          \end{array}   \right)  \
\ee
with each entries being $2 \times 2 $ matrices.
Note that Eqs.(6) and (7) are  non-abelian generalizations of the sine-Gordon
equation and its linear equations. This may be seen easily if we take an
$U(1)$-reduction by setting $g_{1} = g_{2}^{-1} = exp(i \phi \s_3 )$ and $A =
\Ab = 0 $ which satisfies the constraint equation trivially. We may  generalize
further to the non-abelian $A_{r}$-type  Toda (affine Toda) equations if we
take ${\cal G} = diag(g_{1}, \cdots ,g_{r})$ for each $g_{i}$  valued in a Lie
group G, $ \ {\cal A} = diag(A, \cdots , A), \bar{{\cal A}} = diag(\bar{A},
\cdots , \bar{A})$ and  $T = i\k 1 \otimes \L , \Tb = i1 \otimes \bar{\Lambda}$
where  $\Lambda $ is the sum of simple roots of $sl(r)$ (the sum of simple
roots minus the highest root for the affine Toda case) and  $\bar{\Lambda } =
\Lambda ^{t}$.  This, with the obvious generalization of the constraint
equation,  may be  named as  ``the $(A_{r},G)$-Toda (affine Toda) equation" so
that Eq.(6) becomes the affine $(A_{1},SU(2))$-Toda equation. This type of
non-abelian generalizat
of the Toda equation has been first considered by Mikhailov[7] but without the
constraint equation.\footnote{A different type of non-abelian Toda equation has
also been considered in [10]. } In fact, the constraint equation imposes a
highly non-trivial restriction to the model. For example, without the
constraint the abelian limit does not become the sine-Gordon equation but a
pair of coupled two scalar field equations.
Also, it is important to note that the constraint induces the gauge symmetry
which allows different parameterizations of fields depending on particular
choices of gauge fixing. To understand this more clearly, we note that the
potential term added in Eq.(5) is  invariant under the vector gauge
transformation; $g_{1} \rightarrow hg_{1}h^{-1}, \  g_{2} \rightarrow
hg_{2}h^{-1}$ for $h  :  M \rightarrow SU(2)$. It is easy to see that Eqs.(4)
and (6) require $A, \Ab $ to be flat, i.e. $ [ \ \pp + A \ , \ \pb + \Ab \ ] =
0 $ which reflects the vector gauge invariance of the action. For the rest of
the paper, we fix the gauge by setting  $A= \Ab = 0$. In this gauge, the affine
$(A_{1},SU(2))$-Toda equation takes a particularly simple form while the
constraint can not be solved locally.\footnote{In fact, there exists a
different gauge choice which allows solutions of the constraint equation in
terms of local fields. This case will be considered elsewhere.}
\be
\pb(\gi_{1}\pp g_{1}) = \k(\gi_{2}g_{1} - \gi_{1}g_{2}) \ \ ; \  \ \gi_{1}\pp
g_{1}
+ \gi_{2}\pp g_{2} = 0 \ .
\ee

With the linear equation as in Eq.(7), it is now more or less straightforward
to obtain infinite conserved currents of the model. Define $\Phi = \Psi exp(\l
Tz) = \sum_{m = 0}^{\infty}\l^{-m}\Phi_{m}$ with $\Phi_{0} = 1$. If we
parametrize $\Phi_{m}$ by
\be
\Phi_{2m} \equiv \left( \begin{array}{ccc}
p_{2m} & ~ & 0  \\ {}~           & ~   &  ~ \\
 0  & ~ & s_{2m}
          \end{array}   \right)  \ , \
\Phi_{2m+1} \equiv \left( \begin{array}{ccc}
0  & ~ & p_{2m+1}  \\ {}~           & ~   &  ~ \\
 s_{2m+1}  & ~ & 0
          \end{array}   \right)   \ ; \ m \ge 0 \ ,
\ee
the linear equation in each order in $\l $ changes into
\ben
\pb p_{m+1} + i\gi_{1}g_{2}s_{m} = 0 \ & ; & \
\pb s_{m+1} + i\gi_{2}g_{1}p_{m} = 0 \nonumber \\
\pp p_{m} + \gi_{1}\pp g_{1}p_{m} = i\k (p_{m+1}-s_{m+1}) \ & ; & \
\pp s_{m} + \gi_{2}\pp g_{2}s_{m} =  i\k (s_{m+1}-p_{m+1})  \ .
\een
This may be solved iteratively with an initial condition, $p_{0} = s_{0} = 1$.
For example,
\ben
p_{1} &=& {1 \over 2i\k }\gi_{1}\pp g_1 - {1 \over 2i\k }\int dz (\gi_1 \pp g_1
)^2 - {i \over 2}\int d\bar{z} (\gi_1 g_2 + \gi_2 g_1 )  \nonumber \\
s_{1} &=& -{1 \over 2i\k }\gi_{1}\pp g_1 - {1 \over 2i\k }\int dz (\gi_1 \pp
g_1 )^2 - {i \over 2}\int d\bar{z} (\gi_1 g_2 + \gi_2 g_1 ) \ .
\een
 The consistency condition; $\pp\pb p_{m} = \pb\pp p_{m}$ and
$\pp\pb s_{m} = \pb\pp s_{m}$ in Eq.(12) gives rise to two sets of infinite
conserved currents; $\pb J^{(1)}_{m} + \pp \bar{J}^{(1)}_{m+2} = 0; m \ge 0,  $
and  $ \pb J^{(2)}_{m} + \pp \bar{J}^{(2)}_{m+2} = 0 $ where
\ben
 \bar{J}^{(1)}_{m} &=& i\gi_{1}g_{2}s_{m}  \ ; \  J^{(1)}_{m+2}= \pp p_{m+1} =
-\gi_{1}\pp g_{1} p_{m+1} + i\k (s_{m+1} - p_{m+1}) \nonumber \\
 \bar{J}^{(2)}_{m} &=& i\gi_{2}g_{1}p_{m}  \ ; \ J^{(2)}_{m+2} = \pp s_{m+1} =
\gi_{1}\pp g_{1} s_{m+1} + i\k (p_{m+1} - s_{m+1}) \ .
\een
In particular, the energy-momentum conservation; $\pb T_{\pm } + \pp
\Theta_{\pm } = 0$, is given by
\ben
T_{+} &=& i\k(J^{(1)}_{2} + J^{(2)}_{2}) = (\gi_{1}\pp g_{1})^{2} \nonumber \\
\Theta_{+} &=& i \k (\bar{J}^{(1)}_{0}+ \bar{J}^{(2)}_{0}) = -\k (\gi_{1}g_{2}
+ \gi_{2}g_{1})
\een
while the other half of the conservation gives
\ben
T_{-} &=& i\k(J^{(1)}_{2} - J^{(2)}_{2}) = \pp(\gi_{1}\pp g_{1}) \nonumber \\
\Theta_{-} &=& i \k (\bar{J}^{(1)}_{0}- \bar{J}^{(2)}_{0}) = -\k (\gi_{1}g_{2}
- \gi_{2}g_{1})
\een
It is interesting to observe that $T_{-}$ in the abelian limit becomes $T_{-} =
\pp^{2}\phi $ which is precisely the term added to improve the energy-momentum
tensor in the Feigin-Fuchs construction.

Next, we derive $n$-soliton solutions  by using the technique of Riemann
problem with zeros[8].  Take a trivial solution of Eqs.(6) and (7) by
$g_{1}=g_{2} =1 $ and $\Psi = \Psi^{o} = exp(\l Tz - \l^{-1} \Tb \zb )$. Then,
non-trivial solutions can be obtained if we  ``dress" $\Psi_{0}$ by $\Psi
\equiv \Phi \Psi^{o}$, where
$\Phi $ and $\Phi^{-1}$ are matrix functions each possessing $n$ simple poles
with the normalization $\Phi (z,\zb ,\l = \infty ) = 1$.  Here, we assume that
all poles are distinct and  consider soliton solutions for the group $U(2)$
only.\footnote{ For real groups, the requirement of the distinct
pole-assumption should be relaxed.  The group $SU(2)$ and other Lie group cases
as well as the study of properties of solitons will appear in a separate
publication[9].}
The analytic property of $\Phi $ then leads to the linear equation for $\Psi $,
\be
( \pp + U_0 - \l T ) \Psi  = 0 \ , \ (\pb + \li V_1
 )\Psi  = 0 \
\ee
where $U_0$ and $V_1$, given by
\be
U_0 \equiv  -\pp \Phi \Phi^{-1} - \Phi \l T \Phi ^{-1} + \l T
 \ ; \  V_1  \equiv  -\l \pb \Phi \Phi^{-1} +  \Phi \Tb \Phi^{-1} \ ,
\ee
are independent of $\l $. If we identify $U_{0}$ and $V_{1}$ with those of
Eqs.(7)-(9) in the gauge $A=\bar{A} = 0$, we obtain precisely $n$-soliton
solutions. In practice, these identifications can be  simplified a lot if we
make the following reductions;
\newline
{\it i) ${\bf Z}_{2}$-reduction}

 Since the diagonal structure of $U_0$ implies $ Q^{-1} L_{1,2} (\l) Q =
L_{1,2}(-\l) $ for the linear operators in Eq.(7) with $Q = diag(1,-1)$, $\Psi
$ may be reduced by the ${\bf Z}_{2}$-action; $\Psi(\l) = Q \Psi (-\l) Q^{-1} $
or,  $\Phi(\l) = Q \Phi (-\l) Q^{-1} $.
\newline
{\it ii)  Unitarity reduction}

 Unitary of $U$ requires $\J^\dagger (\l) = \J^{-1} (\l^* )$ or
$\Phi ^\dagger (\l) = \Phi ^{-1} (\l^* )$.

\noindent
With reductions imposed, $\Phi$ and $\Phi^{-1}$ may be written as
\be
\Phi = 1+\sum_{s=1}^{n} ({A^s \over \l -\n_{s}} -
{Q A^s Q^{-1} \over \l +\n_{s}}) \ , \
\Phi ^{-1} = 1+\sum_{s=1}^{n} ({B^s \over \l -\m_{s}} -
{Q B^s Q^{-1} \over \l +\m_{s}})
\ee
where $\m_{s} = \n_{s}^{*}$,  and the matrix functions $A^{s}(z,\zb )=
B^{s\dagger }(z,\zb )$ are to be determined. Further determination on $A^{s}$
and  $B^{s}$ comes through the evaluation of residues at $\l =\pm \n_{s},
\pm\m_{s}$ of the equation $\Phi\Phi^{-1}=1$ as well as Eq.(18) which gives
rise to
\be
A^{s} +\sum_{\b=1}^{n} A^s ({B^{\b} \over \n_{s}-\m_{\b}}
                             -{Q B^\b Q^{-1} \over \n_s+\m_\b}) = 0 \ , \
B^{s}+\sum_{\b=1}^{n}({A^{\b} \over \m_{s}-\n_{\b}} -
        {Q A^\b Q^{-1} \over \m_s + \n_\b }) B^s = 0 .
\ee
and
\be
A^s D_{1,2} (\n_s) [1 + \sum_{\a=1}^{n}({B^{\a} \over \n_{s}-\m_{\a}}
 - { Q B^{\a} Q^{-1} \over \n_{s}+\m_{\a}})]
= 0 \ , \
[1 + \sum_{\a=1}^{n}({A^{\a} \over \m_{s}-\n_{\a}}
 - { Q A^{\a} Q^{-1} \over \m_{s}+\n_{\a}})] D_{1,2} (\m_s) B^s = 0
\ee
where $ D_1 = \pp - \l T \ ,  \ D_2 = \pb + \l^{-1} \Tb $. With ans\"{a}tze
$A^{\a}_{ij} = s^{i}_{\a}t^{j}_{\a} \ , \ B^{\a}_{ij} = n^{i}_{\a}m^{j}_{\a}$,
it is easy to show that Eqs.(20) and (21) can be solved in terms of arbitrary
constant vectors $\bar{n}_{\a}$ and $\bar{t}_{\a}= \bar{n}^{\dagger}_{\a}$,
 \be
n_{\a}^{i} = [\J^0 (\m_\a) ]^{ij} \bar{n}^{j}_{\a} \ , \
t^{i}_{\a} = [ \Psi^{o}(\n_{\a})^{-1} ]^{ji} \bar{t}^{j}_{\a} \ ,
\ee
while  $m_{\a}$ and $s_{\a}$ can be  expressed in terms of $t_{\a}$ and
$n_{\a}$ by
\be
m^{l}_{\a} = -{1 \over 2} \sum_{\b=1}^{n} (w^{l})^{-1}_{\a \b} (\n_\b t^l_\b) \
, \
s^{j}_{\a} = {1 \over 2} \sum_{\b=1}^{n}\n_\a n^j_\b (w^{j-1})^{-1}_{\b \a}  \
,
\ee
where $w^{j}_{\a \b} \equiv \sum_{l=1}^{2} \t_{\a \b}
^{1+ (j-l; \ \mbox{mod 2}) } t_{\a}^{l} n_{\b}^{l}$ and
\be
\t^n_{\a \b} \equiv {\n_\a \over 2} ( {1 \over \n_\a - \m_\b } +
  { (-1)^n \over \n_\a + \m_\b } ) \ .
\ee

Having determined $\Phi$ and $\Phi^{-1}$ as above,
 we finally obtain $n$-soliton solutions by evaluating Eq.(18) at $\l=0$ such
that
\be
U_0 = - \pp \Phi \Phi^{-1}, \ \ V_1 = \Phi \Tb \Phi^{-1},
\ee
which gives
$ \Phi^{-1} (\l=0) = {\cal G} = diag(g_1 , g_2 )$ in the gauge $A=\bar{A}
= 0$.
The result is
\be
g_1 = 1 + \sum_{\a=1}^n \sum_{\b=1}^n n^1_\a (w^1)_{\a \b}^{-1} ({\n_\b \over
  \n_\a^*}) t^1_\b \ ; \
g_2 = 1 + \sum_{\a=1}^n \sum_{\b=1}^n n^2_\a (w^2)_{\a \b}^{-1} ({\n_\b \over
  \n_\a^*}) t^2_\b.
\ee
If $n^1_\a $ is invertible,  the $n$-soliton solution can be written  more
compactly with $  N_\a \equiv n^2_\a (n^1_\a)^{-1}$,
\be
g_1 = 1 + \sum_{\a,\b=1}^n {\n_\b^2 - \n_\a^{*2} \over \n_\a^{*2}}
  [ 1 + {\n_\b \over \n_\a^*} {N^{\dagger}}_\b N_\a ]^{-1} \ ; \
g_2 = 1 + \sum_{\a,\b=1}^n {\n_\b^2 - \n_\a^{*2} \over \n_\a^{*2}}
  [ 1 + {\n_\b \over \n_\a^*} {N^{\dagger}}_\b^{-1} N_\a^{-1} ]^{-1}
\ee
where
\be
N_\a = [i \sin \q \bar{n}^{1}_{\a } +
    \cos \q \bar{n^{2}_{\a}} ]
 \times [\cos \q \bar{n}^{1}_{\a } +
    i \sin \q \bar{n^{2}_{\a}} ]^{-1} \ ; \ \theta \equiv \n_\a^* \k z - { 1
\over \n_\a^*} \bar{z} \ .
\ee
It is interesting to check that our soliton solution satisfies the constraint
in Eq.(10). For $n$-soliton solutions,
\be
g_1^{-1} \pp g_1 = -2 i \k \sum_{\a=1}^n s_\a^1 t_\a^2 - 2 i \k \sum_{\a=1}^n
n_\a^2
m_\a^1 \ ; \
g_2^{-1} \pp g_2 = -2 i \k \sum_{\a=1}^n s_\a^2 t_\a^1 - 2 i \k \sum_{\a=1}^n
n_\a^1
m_\a^2.
\ee
With the help of the relation,
$s_\a^1 t_\a^2 = {1 \over 2} \sum_{\a,\b=1}^n ( \n_\a n_\b^1(W^0)^{-1}_{\b \a}
t_\a^2 )
  = - n_\a^1 m_a^2, \ s_\a^2 t_\a^1 =- n_\a^2 m_a^1$,
it is straightfoward to check that $n$-soliton solution indeed satisfies the
constraint. \\

 {\bf $\Phi_{(3,1)}$-deformation } \\

Consider a deformation of the coset model Eq.(2) by adding a  potential term;
\be
I(g_{1},g_{2},A,\Ab ,\b ) = I_{0}(g_{1},g_{2},A,\Ab ) + {\b \over 2\pi }\int
\mbox{Tr}(g_1^{-1} L^a g_1 L^b) \mbox{Tr}(g_2^{-1} M^a g_2 M^b),
\ee
where $\b $ is a coupling constant and the summation is assumed for the
repeated indices $a$ and $b$. We also write $ \{ L^a=\s ^a/2 \} $ and $\{ M^a =
\s ^a /2 \} $($\s^a$ are Pauli matrices) for two sets of generators of $SU(2)$
each corresponding to $g_{1}$ and $g_{2}$ respectively. The potential term in
this case transforms as (triplet, singlet) so that it corresponds to the
$\Phi_{(3,1)}$-deformation. The equations of motion for $g_1 $ and $g_2$ are
\be
[~ \pp + g_1^{-1} \pp g_1 + g_1^{-1} A g_1 , \ \pb + \Ab ~ ]
   +{\b \over 2} \mbox{Tr}(
   g_2^{-1} M^b g_2 M^a) [~ L^a , \ g_1^{-1} L^b g_1 ~ ] = 0 ,
\ee
\be
[~ \pp + g_2^{-1} \pp g_2 + g_2^{-1} A g_2 , \ \pb + \Ab ~ ]
    +{\b \over 2} \mbox{Tr}(
   g_1^{-1} L^b g_1 L^a) [~ M^a , \ g_2^{-1} M^b g_2 ~ ] = 0 ,
\ee
together with the constraint equations,
\ben
&& \mbox{Tr}(L^a ( \pb g_1 g_1^{-1} - g_1 \Ab g_1^{-1} + \Ab) )
 +\mbox{Tr}(M^a ( \pb g_2 g_2^{-1} - g_2 \Ab g_2^{-1} + \Ab) ) = 0,
\nonumber \\
 && \mbox{Tr}(L^a (  -g_1^{-1}\pp g_1 - g_1^{-1} A g_1 + A) )
 +\mbox{Tr}(M^a (  -g_2^{-1} \pp g_2 - g_2^{-1} A g_2 + A) ) = 0.
\een
In order to prove the integrability of the equation, here, unlike the previous
case where we have embedded the product group $SU(2) \times SU(2)$ into the $4
\times 4$ matrix group, we work directly with the product group and denote $g:M
\rightarrow SU(2) \times SU(2)$ by $g = g_{1}g_{2}$.  Note that in these
notations, the potential term  takes a simple form; ${\b \over 2\pi }\int
\mbox{Tr} g^{-1}T g T$ with $T= L^a M^a$. The connections $A, \Ab$, which gauge
both $g_{1}$ and $g_{2}$,  are given by $A= A^{a}L^{a}1_2 + 1_1 A^{a}M^{a},
\Ab = \Ab^{a}L^{a}1_2 + 1_1 \Ab^{a}M^{a}$ where $1_1 , 1_2 $ denote identity
elements. The key observation for the integrability of this product-type model
is that the equations of motion can be merged into a single zero curvature form
with a spectral parameter $\l$ such that
\be
[~ \pp + g^{-1} \pp g  + g^{-1} A g + \b \l T \ , \ \pb + \Ab   +
  \li g^{-1}Tg  ~ ] = 0  \ .
\ee
Using the identity $L^a L^b = {1 \over 4} \d^{ab} +{i \over 2} \e ^{abc} L^c$
for $SU(2)$, the equivalence between Eq.(34) and Eqs.(31) and (32) can be shown
directly  by evaluating  the coefficient terms of $L^a 1_2$ and $1_1 M^a $ in
Eq.(34) which gives rise to Eq.(31) and Eq.(32) whereas the coefficient term of
$L^a M^b$ vanishes identically.

The zero curvature expression for the model  again leads to  the infinite
conserved currents through the same procedure as in the generalized Toda case.
By making use of the constraint equation, we first define $B$ field by
\be
g^{-1} \pp g + g^{-1} A g = A^a (L^a 1_2 + 1_1 M^a ) + B^a (L^a 1_2 - 1_1 M^a )
\ee
and reexpress the linear equation in terms of $\Phi \equiv \Psi H^{-1} exp( \l
\b Tz)  \equiv \sum_{m = 0}^{\infty }\l^{-m}\Phi_{m}, \ \Phi_{0}=1 $ where $H$
solves the flat connection resulting from Eqs.(31)-(33);
$A= H \pp H^{-1}, \Ab = H \pb H^{-1}$,
\be
\pp \Phi_m + A^a [L^a 1_2 + 1_1 M^a , \Phi_m] + B^a (L^a 1_2 - 1_1 M^a ) \Phi_m
  + \b [ L^b M^b, \Phi_{m+1}] = 0,
\ee
\be
\pb \Phi_m + \Ab^a [L^a 1_2 + 1_1 M^a , \Phi_m]
 + (g_1^{-1} L^b g_1) (g_2^{-1} M^b g_2) \Phi_{m-1} = 0.
\ee
This may be solved iteratively with a parametriztion of  $\Phi_m$ by
\be
\Phi_m \equiv \a_m 1_1 1_2 + \b_m ^a (L^a 1_2 + 1_1 M^a ) + \g_m^a (L^a 1_2 -
1_1 M^a )
 + \d_m^a \e ^{abc} L^b M^c
 + \h_m ^{ab} (L^a M^b + L^b M^a)
\ee
where $\h_m ^{ab} = \h_m ^{ba}$.  In the gauge $A^a = \Ab^a = 0$, the iterative
solutions for each component are;
\ben
\g_m^a &=& {i \over {2 \b}} ( \pp \d_{m-1} ^a + i \e ^{abc} B^b \b_{m-1} ^c ),
 \nonumber \\
\d_m^a &=& - {2 i\over \b}  ( \pp \g_{m-1}^a + B^a \a_{m-1}
 + i \e^{abc} B^b \b _{m-1}^c -{1 \over 2} B^b \h_{m-1}^{ab} )
\nonumber \\
\a_m &=& - {1 \over 2} \int B^a \g_m^a dz - {1 \over 16} \int (
 \e^{abc} \d_{m-1}^a D_{(A)}^{bc} + 2 D_{(S)}^{ab} \h_{m-1}^{ab} ) d \bar z
\nonumber \\
\b_m^a &=& -\e^{abc} \int ({i \over 2} B^b \g_m^c - {1 \over 4} B^b \d_m^c ) dz
  -{1 \over 4} \int (D_{(S)}^{ab} \b_{m-1}^b + D_{(A)}^{ab} \g_{m-1}^b +
  {i \over 2} D_{(A)}^{ab} \d_{m-1}^b ) d \bar z
\nonumber \\
\h_m^{ab}&=&{1 \over 2} \int ( B^a \g_m^b + {i \over 2} B^a \d_m ^b -
   {i \over 2} B \cdot \d_m \d^{ab}  ) dz
- {1 \over 4} \int (D_{(S)}^{ab} \a_{m-1} + i \e^{acd} D_{(S)}^{bc} \b_{m-1}^d
\nonumber  \\
& &+i \e^{acd} D_{(A)}^{bc} \g_{m-1}^d + {1 \over 4} \e^{acd} D_{(A)}^{cd}
 \d_{m-1}^b -{1 \over 2} D_{(S)}^{cd} \e^{ace} \e^{bdf} \h_{m-1}^{ef}
) d \bar z  + (a \Leftrightarrow b)
\een
where
\be
2 D_{(S)}^{ab} \equiv { D^{ab}(g_2^{-1} g_1) + D^{ba} (g_2^{-1} g_1 ) } ,
D^{ab} (g_1) = 2 \mbox{Tr} (g_1^{-1} L^a g_1 L^b)\ , \ 2 D_{(A)}^{ab} \equiv {
D^{ab}(g_2^{-1} g_1) - D^{ba} (g_2^{-1} g_1 ) }.
\ee
With an initial condition $\Phi_{0} = 1 $, the first iterative solution, for
example, is
\be
\a_1 = \b_1^a = \g_1^a = 0,\
\d_1^a = - {2 i \over \b } B^a,\  \h_1^{ab}
  = \int(B^a B^b + \d^{ab} B \cdot B) dz
 -{1 \over 2} \int D_{(S)}^{ab} d \bar z.
\ee
Also, the consistency condition; $\pp \pb \Phi_{m} = \pb \pp \Phi_{m} $ yields
a large set of infinite conservation laws  in components through Eqs.(36)-(39)
whose explicit form can be easily written down. Here, however we simply report
the first non-trivial conservation law,
\be
\b \pp D_{(S)}^{ab} + 2 \pb ( B^a B^b - B \cdot B \d^{ab} ) = 0
\ee
which implies the energy-momentum conservation,
\be
\b \pp D_{(S)}^{aa} - 4 \pb (  B \cdot B  ) = 0
\ee
with $B^a = (g_1^{-1} \pp g_1 )^a = - (g_2^{-1} \pp g_2 )^a$.

In this Letter, we have introduced integrable deformations of coset models by
adding potential terms which correspond to $\Phi_{(2,1)}$ and $\Phi_{(3,1)}$
operators. We have demonstrated the integrability by deriving explicitly
infinite conservation laws in each case. The deformation by the operator
$\Phi_{(3,1)}$ has not been taken seriously in previous works since the
operator becomes irrelevant for $c < 1$. However, for $c=1$, it becomes
marginal and since our work proves the classical integrability of the deformed
coset model with coset
$(SU(2)_N \times SU(2)_N)/SU(2)_{2N}$ for all $N$, it would be interesting
to understand  physical implications of this type of deformation as well as
those of the Toda-type deformation.

\newpage

{\bf ACKNOWLEDGEMENT}
\vglue .2in
 We would like to thank  D.Kim, I.Bakas, T.Hollowood, J.Evans for discussion
and B.K.Chung and S.Nam for help.  Q.P. would like to thank the Theory Division
at CERN  for its hospitality while this work was completed. This work was
supported in part by Research Fund of Kyunghee University, by the program of
Basic Science Research, Ministry of Education BSRI-94-2442, and by Korea
Science and Engineering Foundation.
\vglue .2in

\def\Item{\par\hang\textindent}

{\bf REFERENCES }
\vglue .2in

\Item {[1]}  A.Zamolodchikov, Sov.Phys.JETP Lett. {\bf 46} (1987); Int.J.Mod.
Phys. {\bf A3} (1988) 743.
\Item {[2]} T.Eguchi and S.Yang, Phys.Lett.{\bf B224} (1989) 373.
\Item {[3]} T.Hollowood and P.Mansfield, Phys.Lett.{\bf B226} (1989) 73.
\Item {[4]} D.Karabali, Q-H.Park, H.J.Schnitzer and Z.Yang, Phys.Lett.{\bf
B216} (1989) 307; D.Karabali and H.J.Schnitzer, Nucl.Phys.{\bf B329} (1990)
649;
K.Gawedski and A.Kupiainen, Phys.Lett. {\bf B215} (1988) 119, Nucl.Phys.{\bf
B320} (1989) 625.
\Item {[5]} E.Witten, Commun.Math.Phys.{\bf 92} (1984) 455.
\Item {[6]} Q-H.Park, Phys.Lett.{\bf B328} (1994) 329.
\Item {[7]} A.V.Mikhailov, Physica {\bf 3D} (1981) 73.
\Item {[8]} V.E.Zakharov, S.V.Manakov, S.P.Novikov and L.P.Pitaievski, Theory
of Solitons, Moscow, Nauka 1980 (Russian); English transl.:New York,  Plenum
1984.
\Item {[9]} Q-H.Park and H.J.Shin, in preparation.
\Item {[10]} A.N.Leznov and M.V.Saveliev, Commun.Math.Phys.{\bf 89} (1983) 59.
\Item {[11]} I.Bakas, Int.J.Mod.Phys. {\bf A9} (1994) 3443.
\end{document}